\documentclass[preprint]{aastex}

\begin{document}

\title{M31 Globular Clusters in the HST Archive: II. Structural Parameters%
\footnote{Based on observations made with 
the NASA/ESA Hubble Space Telescope,
obtained from the data archive at Space Telescope Science Institute. 
STScI is operated by the Association of Universities
for Research in Astronomy, Inc. under NASA contract NAS 5-26555.}
}

\author{Pauline Barmby\altaffilmark{2}}
\affil{Harvard-Smithsonian Center for Astrophysics, 60 Garden St., Cambridge, MA 02138} 
\email{pbarmby@cfa.harvard.edu}
\author{Stephen Holland}
\affil{Department of Physics, University of Notre Dame, Notre Dame, IN 46556-5670}
\email{sholland@nd.edu}
\author{John P. Huchra}
\affil{Harvard-Smithsonian Center for Astrophysics, 60 Garden St., Cambridge, MA 02138} 
\email{huchra@cfa.harvard.edu}
\altaffiltext{2}{Guest User, Canadian Astronomy Data Centre, 
which is operated by the Herzberg Institute of Astrophysics, National Research Council of Canada.}

\shortauthors{Barmby, Holland, \& Huchra}
\shorttitle{Structure of M31 globulars}

\begin{abstract}

We analyze the structural parameters of the largest-available sample of spatially 
resolved extragalactic globular clusters. The images of M31 GCs were found in
a search of HST archival data, described in a companion paper.
We measure the ellipticities and position angles of the clusters and conclude
that the ellipticities are consistent with being caused by rotation. 
We find that most clusters' surface brightness distributions are
well-fit by two-dimensional single-mass Michie-King models. A few clusters
show possible power-law distributions characteristic of core-collapse, but the
spatial resolution is not high enough to make definitive claims.
As has been found for other galaxies, the metal-rich clusters
are slightly smaller than the metal-poor clusters. 
There are strong correlations between structural properties of M31
GCs, as for Milky Way clusters, and the two populations are
located close to the same `fundamental plane' in parameter space.

\end{abstract}

\keywords{galaxies: individual (M31) -- galaxies: star clusters}

\section{Introduction}

Globular clusters' structures 
yield important information about  their dynamical states and also about
the environmental effects of the parent galaxy's tidal field. The 
core radii and ellipticities of Galactic globulars are known to
vary with position \citep{vdb94,ws87}, and the mean structural
parameters of clusters also vary between galaxies. For example, the
LMC's globular clusters are, on average, much more 
flattened than those of the Milky Way \citep{vdb83}.
These findings have led to many suggestions and theories about the
histories of GCSs and their parent galaxies \citep[e.g.,][]{vdb00},
but these conclusions would be much stronger if structural
parameters were accurately measured for clusters in more galaxies.

Globular clusters in Local Group galaxies are particularly
valuable for such a comparison. They are distant enough that their 
integrated properties can be easily derived, but
near enough that their individual stars can also be
resolved. M31 has the Local Group's largest
globular cluster population, so it is a natural starting place for
studies of extragalactic globular clusters.
The M31 GCS is known to be similar to the Milky Way's in numerous
ways \citep[e.g., metallicity and spatial distributions;][]{b00}, but has
some important differences as well 
\citep*[chemical composition and possibly age and luminosity 
distribution:][]{bh91,bhb01}. While many
properties of the M31 globular clusters can be elicited with
ground-based imaging and spectroscopy, high spatial
resolution data from space-based telescopes such as HST 
are required to study their spatial structures and stellar populations.
The angular sizes of M31 globular clusters are comparable
to the sizes of ground-based seeing disks: \citet{s79,s81}
and \citet{hol_phd}
showed that seeing can lead to substantial errors in the
derived structural parameters of galaxies with similar sizes.
The first HST measurements \citep{ben93,fp94} showed that the
ground-based measurements of core and half-light radii
were systematically overestimated. To date, post-repair HST measurements
of M31 GC parameters (\citealp{cmd4,gri96}; \citealp*{hfr97}) have 
included only a total of seven objects. 

In a companion paper \citep[hereafter Paper I]{hst_1}, we 
described the use of HST/WFPC2 images to
identify globular clusters and candidates in M31 and thus
examine the quality of existing cluster catalogs. That paper
contains the details of the HST Archive search and data reduction procedures.
In this paper, we derive structural parameters for `good' (category A and B) 
cluster candidates found in our survey. The objects studied here do not
constitute a complete or magnitude-limited sample of globular clusters
in M31, and only the clusters which were not specific HST targets
can be considered to be a random sample. Some of our new unconfirmed
cluster candidates may not be M31 globulars at all
Several of the clusters in our sample --- the targets of 
HST program GO-6699 --- may actually belong to M31's companion galaxy NGC~205.
We retain these objects in our sample 
except when computing correlations between cluster properties and
location in the galaxy.
We have studied a much larger sample of M31 globular clusters
than has been previously available, and this allows us
to explore correlations of structural parameters with other
properties such as galactocentric distance, and to determine
mean structural properties for comparison with other galaxies.

\section{Surface Brightness Distributions}

\subsection{Color gradients}

Aperture photometry in multiple filters can be used to search for 
radial color gradients in GCs. Since there is no evidence for internal 
extinction in GCs, radial color variations are thought to indicate variations
in the cluster stellar population. 
\citet{dj91a,dj91b} studied color variations in 12 Milky Way globulars 
over linear scales of 0.04--4~pc and found the core-collapsed 
clusters to be bluer in the center, either because of
an excess of blue straggler stars or a deficit of red giants. 
The color gradients in the core-collapsed clusters were
0.1--0.3~mag per decade in radius; non-collapsed clusters did not show 
any color gradients. The measurement of color gradients in globular clusters 
is subject to a number of systematic uncertainties, including
flat-fielding, contamination from background objects and from M31, absolute
centering, centering differences between filters, and stellar population
sampling. Using color gradients to detect core-collapsed clusters
in M31 is an attractive idea, since it might require lower spatial resolution
than a detection from the surface brightness profile. Because of the
effects listed above, we believe that such a detection would require
a much more detailed study than is presented here, including
photometry of individual stars and a careful treatment of incompleteness.
This examination of integrated color profiles is intended to serve
as a preview for a more detailed study.

\begin{figure}
\includegraphics*[scale=0.7]{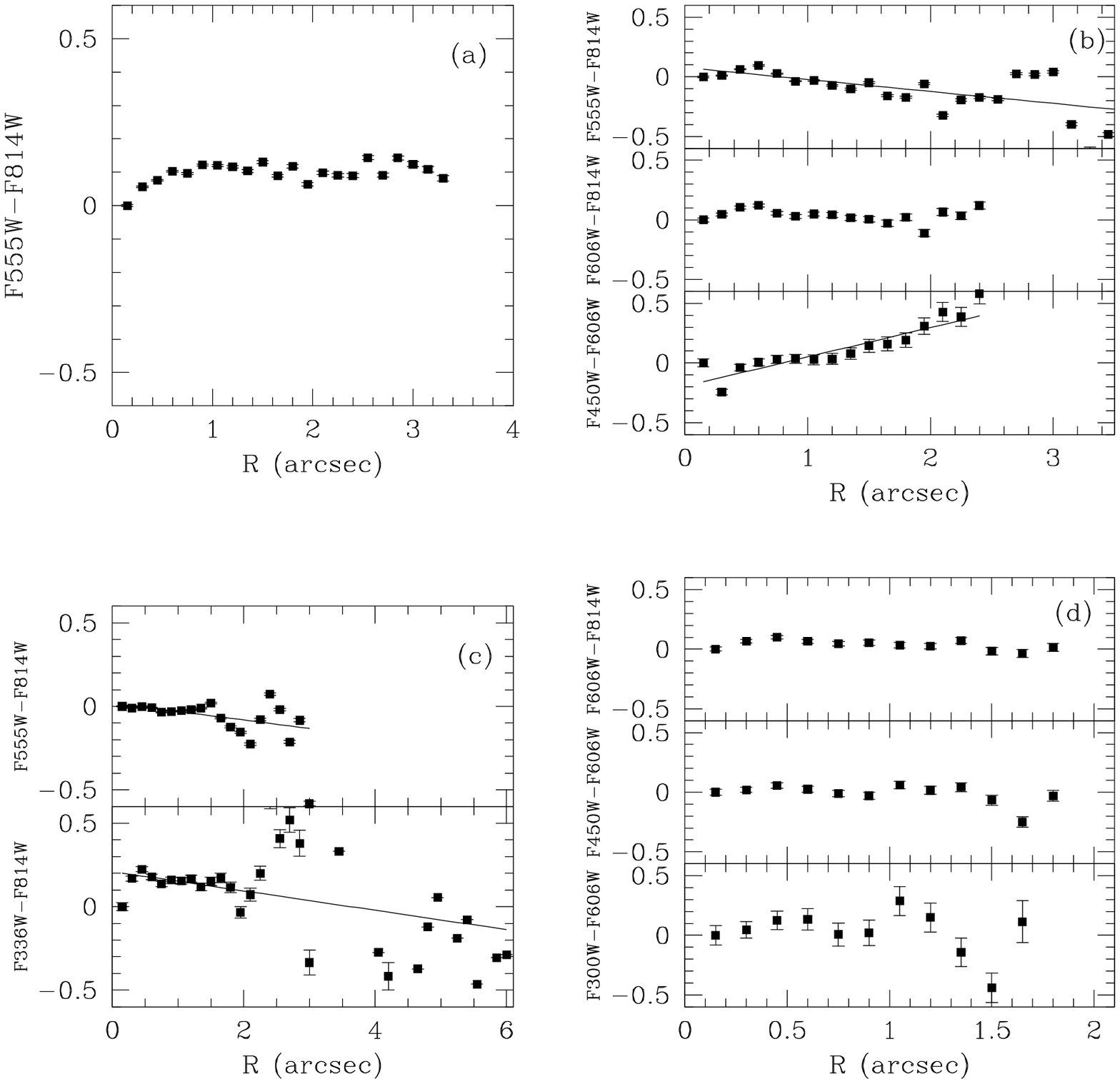}
\caption
{Sample color profiles of M31 globular clusters:
(a) 240--302 (b) 109--170 (c) 127--185  (d) 076--138.
The colors plotted are relative to the color measured in the central
aperture. Note that the horizontal axis scales are not identical. Solid lines are
the least-squares fits to the color profiles of clusters with gradients.
\label{fig-colorprof}}
\end{figure}

The aperture photometry described in Paper~I was used to measure the
color profiles. The color in each annulus was computed as 
\begin{equation}
C_i = C_0 -2.5\log\left(\frac{f_1(r_i)-f_1(r_{i-1})}{f_2(r_i)-f_2(r_{i-1})}\right)
\end{equation}
where $f_1$ and $f_2$ refer to the fluxes measured in the two different filters.
All of the color profiles are referenced to the color in the central aperture $C_0$. 
The color profiles had a wide variety of appearances: we show a sample in 
Figure~\ref{fig-colorprof}. As expected, the profile appearance was much
more `ragged' for shorter exposure times and bluer filters, which
yielded many fewer counts and hence much more uncertain fluxes.
Bright, rich clusters tended to have smooth
color profiles, while the looser clusters had much more variable profiles,
as might be expected from population sampling statistics. 
To illustrate this, we used the method of \citet{ren98} to estimate the projected
number density of red giants inside the clusters' half-light radii 
in the absence of population gradients. The values ranged from
$\sim 200$ stars~arcsec$^{-2}$ for typical bright clusters to
$\sim 2$ stars~arcsec$^{-2}$ for the faintest and loosest clusters.

We checked the profiles for significant
color gradients using a Monte Carlo technique, resampling each profile $10^4$ times
by bootstrapping and computing the weighted least-squares fit to a straight line. If the 
absolute value of the computed slope for the real dataset was in the 95$^{\rm th}$ 
percentile of the distribution of the absolute value of the slope 
for the bootstrapped datasets, we considered it to be significant. 
There were a total of 18 significant color gradients, with two objects having gradients
in more than one color. The gradients are comparable in size to those obtained by 
\citet{dj91a} for Milky Way globulars. Strangely, the two gradients for 109--170 (see 
Figure~\ref{fig-colorprof}b) are in opposite
directions: the center of the cluster is redder in F450$-$F606W and bluer in F555W$-$F814W. 
The F450W/F606W and F555W/F814W image pairs were obtained at different times, so
the different color gradients might be due to a bright variable star in the cluster
or to cosmic ray contamination in one image pair.
H$\alpha$ emission could also have contaminated the F606W image. 
Overall, two thirds of the slopes are positive (the center is bluer, as found for Milky Way 
core-collapse objects) and one third are negative (as might be expected for objects
which are actually background galaxies).

There is some previous work on M31 GC color gradients: 
\citet{hfr97} found gradients of $-0.028$ and $-0.090$~magnitudes arcsec$^{-1}$ 
for 240--302 and 379--312, respectively. 
We find no significant gradient for 240--302 (shown in Figure~\ref{fig-colorprof}), and 
a smaller gradient, $-0.02$~mag~arcsec$^{-1}$, 
for 379--312. Since we use the same observational data as \citet{hfr97}, 
the discrepancy may be related to the fitting or background subtraction methods. 
\citet{gri96} found no significant color gradients for 006--058, 045--108,
343--105, and 358--319; we concur with their results for the first three clusters
and find a small gradient (0.02 mag~arcsec$^{-1}$) for 358--319. 
The small sizes of the well-determined gradients and the many possible 
systematics suggests that definitive measurements of population gradients in 
M31 GC using integrated colors requires a more detailed study, and
possibly better data, than are available here.

\subsection{Shape and Structural Parameters}

To measure clusters' shapes, we used the IRAF task {\sc ellipse} to fit 
elliptical isophotes to the background-subtracted images. 
Isophotes were fit over a range of semi-major axes
spaced logarithmically from 0.2\arcsec\ to 5.0\arcsec\ or the largest
measurable size. {\sc ellipse} could not be made to converge for 9
sparse clusters. This was not unexpected, since the algorithm
was designed for galaxies and expects the surface brightness distribution to 
be smooth and monotonically decreasing outward.
We were able to estimate shape parameters for about a third of these
objects by resampling the images to a resolution of 0.2\arcsec/pixel
and running {\sc ellipse} on the `blurred' images. This procedure did not
work for the remainder of the images, and objects with no shape
measurements are noted in Table~\ref{tbl-param} with a default zero
ellipticity and position angle.

\begin{figure}
\includegraphics*[scale=0.7]{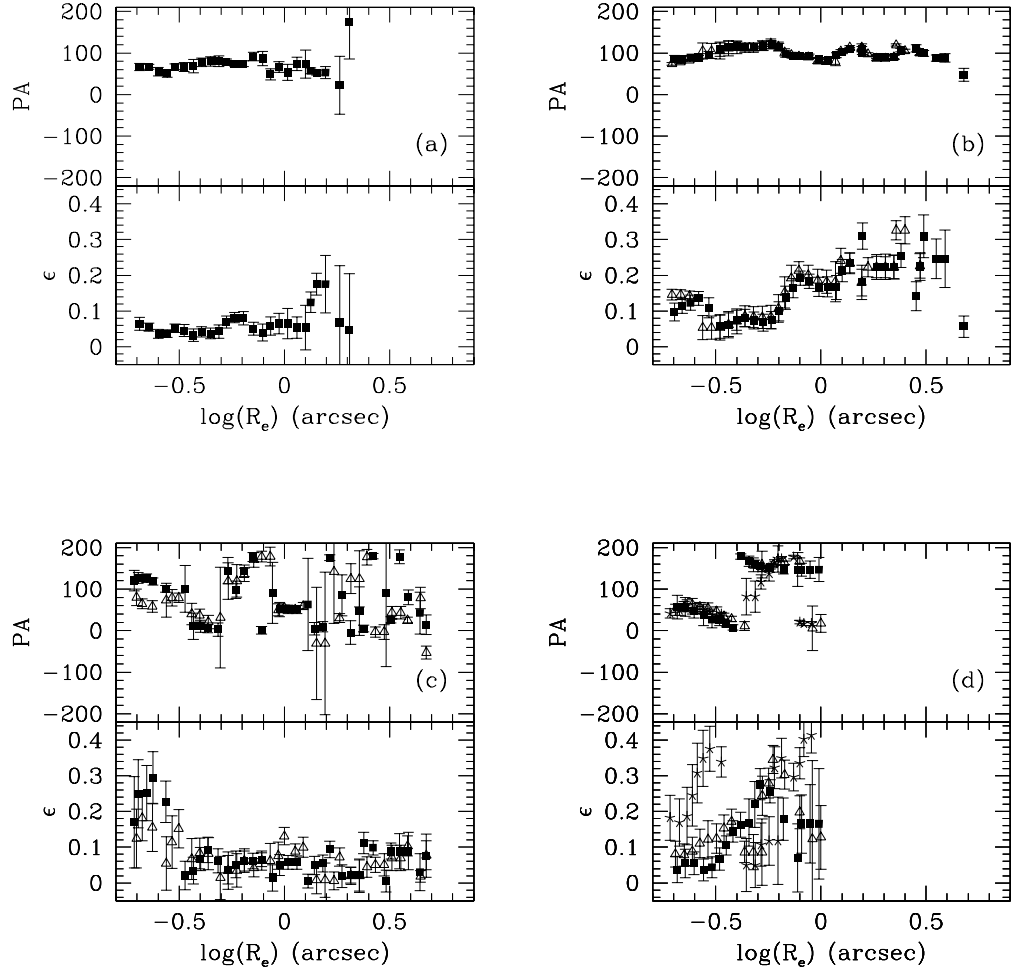}
\caption
{Sample ellipticity and position angle profiles for M31 GCs:
(a) 153--000 (F300W) (b) 240--302 (F555W, F814W) (c) 338--076 (F555W, F814W) (d) NB39 (F300W, F555W, F814W)
\label{fig-ellip_prof}}
\end{figure}

Figure~\ref{fig-ellip_prof} shows some sample ellipticity 
($\epsilon=1-b/a$) and position 
angle profiles, plotted as a function of the effective radius 
($R_e=\sqrt{ab} =a \sqrt{1-\epsilon}$)
to allow for simple comparison of objects with different ellipticities.
In most cases, the
ellipticities and position angles measured on images in different
filters track well together, as we would expect.
The measured position angles are occasionally
wildly varying, often when the ellipticities are close to
zero. This is likely an artifact of the {\sc ellipse} algorithm,
which diverges as the ellipticity approaches zero \citep{jed87}. 
For these objects, we report ellipticities and position angles 
of zero in Table~\ref{tbl-param}.
We averaged the {\sc ellipse} output over the isophotal semi-major axes
to determine the overall ellipticity, position angle, and central position 
for each of the cluster images. We further averaged over the different filters to
compute the ellipticities and position angles
given in Table~\ref{tbl-param}.

Our measured ellipticities generally agree quite well with those of \citet*{ssg96}, 
the most comprehensive published dataset. The median difference
between two two sets of measurements is $0.024\pm0.009$.
Our measured position angles do not agree particularly well
with those measured by \citet{ssg96}, although
the agreement is better for more elliptical objects, 
for which the position angle can be more precisely determined.
There are numerous possible sources of systematic differences in 
ellipticity and position angle measurements, including isophote centering,
size of the semi-major axis steps, the adopted algorithm,
the image bandpass, and seeing and guiding errors (which could
have significant effects on the ground-based results).
A comparison of position angles measured by \citet{ssg96} and \citet{lup89}
shows that the measurements in these two works also agree rather poorly.
As most of the ellipticities are small, precise measurement of
position angles is not critical to our surface brightness modeling
and the discrepancies are not a serious concern. The median
ellipticity of the M31 clusters is $0.11\pm0.01$, in 
reasonably good agreement with the measurements of 
\citet{ssg96} ($0.09\pm0.04$) and \cite{lup89} ($0.08\pm0.02$).
The clusters' position angles show no tendency
to align with either the major or minor axes of M31, or with the 
local direction toward the center of the galaxy.
This indicates that tidal forces are not responsible for the
M31 GCs' ellipticities; \citet{ws87} came to the same conclusion
for the Milky Way clusters.

We next fit single-mass, elliptical Michie-King \citep{mic63,kin66}
(hereafter simply `King models') to the cluster images. 
As usual, we parameterize the models with the scale radius%
\footnote{\citet{bm98} point out that $r_0$ is usually called the
core radius and denoted $r_c$, but $r_0$ as defined by \citet{kin66}
is approximately the same as $r_c$ (the half-intensity radius) 
only for concentrated clusters.}
$r_0$, the concentration $c=\log(r_t/r_0)$ ($r_t$, the tidal radius,
is where the projected cluster density drops to zero), and ${\mu}(0)$,
the central surface brightness.
We used the program {\sc km2dfit} written by S. Holland
\citep*[described in][]{hhc99} to do the model fits. Although {\sc km2dfit}
can also fit for the ellipticity, position angle 
and central position, increasing the number of 
parameters greatly increases the execution time so we used
the {\sc ellipse} values for these parameters instead.
For most objects, we fit the models to the clusters over sub-images 
12.8\arcsec\ in size. 
Some objects near the edges of the WFPC2 chips had to be fit
in smaller sub-images, and a few were so close to the chip edge that
they could not be fit at all (these are marked in Table~\ref{tbl-param}).

The models were convolved with the appropriate
WFPC2 PSF before being compared to the data; this greatly increased the
execution time but should result in more accurate parameters for the
smaller objects. We tried fitting models without
PSF convolution, and found that the resulting
scale radii were systematically larger (by $0.076\pm0.013\arcsec=0.3$~pc) and
the concentrations smaller (by $0.09\pm0.02$) than for the convolved models. 
This shows that the size of the PSF cannot be ignored, even with 
HST resolution. To check the effect of pixel size (i.e., whether a
cluster  was imaged on the PC or one of the WFC chips), we 
selected PC images of 10 clusters with a range of structural parameters, 
rebinned them to the lower WFC resolution, and re-fit models to the 
rebinned images. The differences between the
PC and WFC models were small: median offsets were  
$0.01\pm0.04\arcsec=0.04\pm0.16$~pc in $r_0$, and $0.04\pm0.04$ in $c$.

Measurement of the same object imaged in more than one filter 
generally gave quite similar results; the median absolute differences 
in recovered parameters
were $0.13\pm0.04$ in $c$ and $0.04\pm0.02\arcsec=0.15\pm0.08$~pc in $r_0$.
These values provide reasonable estimates of the systematic uncertainties
in $c$ and $r_0$. The tidal radii are more uncertain since they
depends on both of these parameters: error propagation for median
values of $c$ and $r_0$ yields an estimate of $\Delta r_t \approx 2$\arcsec.
The situation was similar for objects imaged in more than two filters,
although fits in the F300W and F336W were often discrepant from others.
The poorer signal-to-noise in these filters and/or a different spatial
distribution of the horizontal-branch stars (which emit most of the UV light)
are the likely causes of the discrepancy.
Table~\ref{tbl-param} gives average King model parameters $r_0$, $c$ and $r_t$
for each cluster, and the central surface brightness in the $V$-band
(or another filter if $V$ was unavailable).
The central surface brightness is determined by transforming
the model central intensity in counts per pixel
to magnitudes per square arcsecond using the same calibration as
in Paper~I. 

It would be useful to know if there were any systematic effect of
exposure time on the model-fitting results. Once could imagine that
longer exposure times $t$ relative to a cluster's integrated magnitude
might allow better detection of faint stars at the
edge of the cluster, and hence yield larger $r_t$ and $c$ values.
Unfortunately it is not possible to directly examine the relationship
between the number of photons in a cluster's image 
$N_p \propto t \times 10^{-0.4V}$ and the derived King-model parameters.
$N_p$ is strongly related to cluster integrated magnitude --- brighter clusters 
emit more flux {\em and\/}, in our HST sample, generally have larger $t$ --- which
is known to correlate with $c$ for Milky Way clusters
(we will show below that the same correlation holds for M31 clusters).
Instead, we sorted the clusters into 18 pairs with nearly the same
$V$ magnitudes ($\Delta V<0.1$~mag) and values of $N_p$ differing by more than
a factor of 1.5. The clusters with larger $N_p$ had larger values of $r_0$
in 11 of 18 cases, and larger values of $c$ in 10 of 18 cases.
We conclude that there does not appear to be a systematic
difference in the measured cluster parameters with $N_p$.

\begin{figure}
\includegraphics*[scale=0.7]{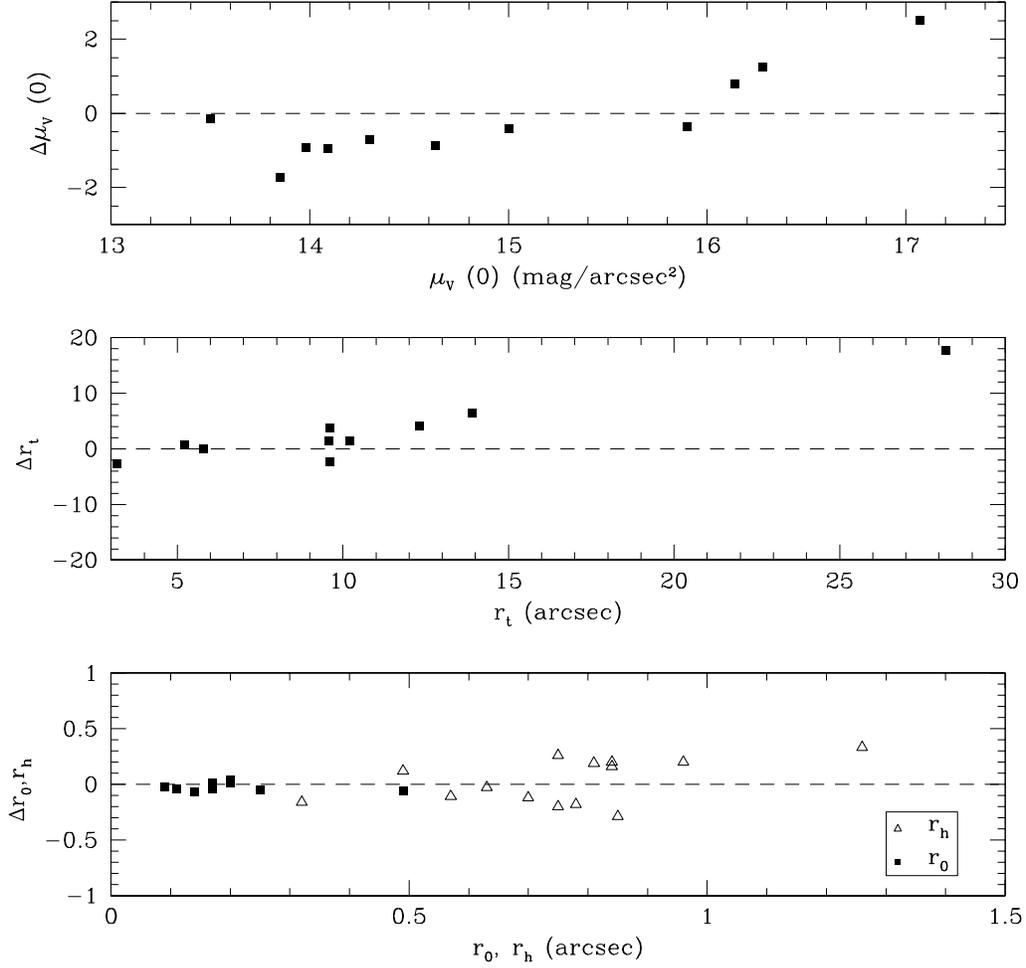}
\caption
{Comparison of our measurements of structural parameters with those of
previous authors. The horizontal axis is the published measurement; vertical
axis is (published $-$ our) measurements.
\label{fig-structcomp}}
\end{figure}

In Figure~\ref{fig-structcomp}, we compare our measurements of structural
parameters with previous HST and ground-based measurements.
The agreement is good for $r_0$ and $r_h$ and rather poor for ${\mu}_V(0)$ and $r_t$.
The poor agreement for the central surface brightness is likely due
to cluster flux being smeared out by the PSF in the previous
measurements. \citet{fp94} used deconvolved (pre-COSTAR) Faint Object 
Camera images, and \citet{dp90} used ground-based images with PSF deconvolution. 
The poor agreement in $r_t$ may be attributable to the ground-based measurements
by \citet{cf91}, which are highly uncertain for individual clusters. 
The large bright cluster 000--001 (also known as G1 or Mayall~II)
has been previously studied by \citet{cmd4} and \citet{mey01}.
The structural parameters for all three works are given in Table~\ref{tbl-g1};
al agree fairly well on the values of $r_0$ and ${\mu}_V(0)$
but disagree but about a factor of 5 on $r_h$ and $r_t$.
Different methods used by the different groups may be the
causes of the disagreement. The $r_t$ value of \citet{cmd4}
is based on a detection of the tidal cutoff in the
surface brightness profile, while ours is based on the $c$ 
measured from the overall shape of the profile. Because most
of the weight in our fit comes from the bright inner regions
of the cluster, small variations from a pure single-mass King profile
will bias our $r_t$ away from the true value. \citet{mey01}
fit multi-mass (instead of single-mass) King models to the one-dimensional 
surface brightness profile, although they do not correct for the PSF.
\citet{gg79} suggest that it is quite reasonable for 
$r_t$ in a multi-mass model to differ by a factor of 2 from
that in a single-mass model. G1 is an interesting cluster for
many reasons, and a more detailed investigation
its surface brightness distribution than is carried out here
could be useful.

\begin{figure}
\includegraphics*[scale=0.7]{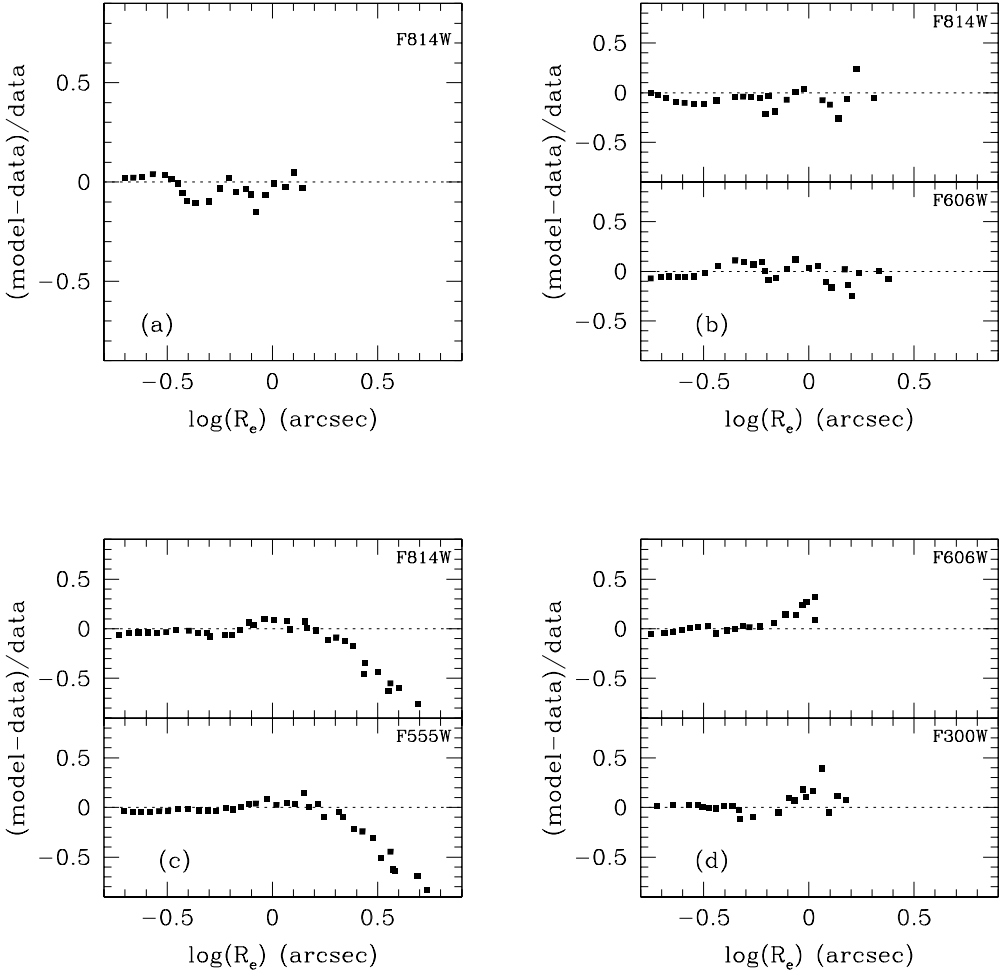}
\caption
{Difference between modeled and measured
surface brightness profiles for a sample of M31 GCs: (a) 020D--089
(b) 160--214 (c) 006--058 (d) 374--306. 006--058 may have extra-tidal stars. 
\label{fig-sbprof}}
\end{figure}

A final step in checking the modeling results is the examination 
the differences between modeled and measured surface brightness 
profiles. Figure~\ref{fig-sbprof} shows a sample of 
these residuals, which are generally less than 10\%.
The figure also demonstrates the different physical sizes of
the clusters; the points stop at the radius where {\sc ellipse}
can no longer fit the isophotes because of poor signal-to-noise.
An important question to be addressed by examining the profiles
is whether there is evidence for systematic departures of the data from the model
profiles. Departures at large radii can indicate the presence of
extra-tidal stars, while departures at small radii can indicate the
presence of the core-collapse phenomenon. Both effects have been
claimed in M31 GCs, by \citet{hfr97}, \citet{gri96}, and \citet{fp94}. 
Examining the profiles,
we find evidence that the following clusters appear to have extra-tidal stars:
006--058, 058--119, 110--172, 240--302, 358--219, and 379--312. 
We are in agreement with the results of
\citet{hfr97} and \citet{gri96} for all objects except 343--105,
which \citeauthor{gri96} find to have extra-tidal stars and we do not.
A potential problem with detections of excess flux at large radius
is the uncertain effects of background subtraction.
We tried to account for this in our model-fitting by allowing the
background level to vary even though it should have been set to zero by our
subtraction procedure. 

Detecting core-collapsed clusters is difficult: even in the
Milky Way, detections of core-collapse in GC surface brightness profiles 
came many years after the phenomenon was first predicted 
\citep[see, e.g.,][]{dk84}. Core-collapsed globular clusters are distinguished 
from `King-model' clusters by the fact that their surface brightness profiles 
are better fit by a power law. To check for core collapse in M31 GCs, we fit power-laws 
to the {\sc ellipse} surface brightness profiles and compared the RMS deviation between 
the power-law model profiles and the data to that between King model profiles and data. 
As expected, most of the clusters were better fit by King models than by power laws. 
The profiles of a few objects, mostly those for which the best-fit King models had 
large values of $c$, were fit as well as or better than King models by power laws.
These may be core-collapsed clusters; we show their profile residuals in 
Figure~\ref{fig-corecoll} and mark them in Table~\ref{tbl-param}.
\citet{ben93} and \citet{gri96} both suggested that 343--105 
showed signs of core collapse in its profile, while we find that
the King model is formally a slightly better fit than the power law for this object.
The two previous works deconvolved the observed profile from the PSF 
before fitting a power law, which may be why they measured
a slightly different profile. However, as Figure~\ref{fig-corecoll} shows, 
there is very little difference between the two models, and we 
believe it is very difficult to use the existing data to differentiate between the profile of
a high-concentration King model with a small scale radius (for 343--105 we measured
$r_0=0.42$~pc) and a power-law one with existing data.

\section{M31 and Milky Way Globular Cluster Comparisons and Correlations}

The structural parameters of globular clusters are the result of
both their current and past dynamical conditions. It is therefore of interest to compare
the measurements of M31 globular clusters to those of clusters in
other galaxies, primarily the Milky Way, 
and to search for correlations among their properties.
We used the June 1999 version of the \citet{h96} catalog
as our source of Milky Way cluster properties%
\footnote{Note that the
central surface brightness measurements reported in the Harris catalog are not
corrected for extinction, although the absolute magnitudes are.}, 
supplemented by the \citet{ws87} data on cluster ellipticities. 

\subsection{King Model Parameters}

M31 and Milky Way globular cluster structural parameters are shown in Figure~\ref{coreplots}.
While the two galaxies' GCs follow essentially the same trends, the M31 clusters 
cover a much smaller range of sizes and central surface brightnesses
than the Milky Way clusters.
The largest scale radius for a Milky Way cluster is 20.2~pc (Pal~14); the largest
scale radius we measured for a confirmed M31 GC is 6.1~pc (468--000).
Comparing M31 clusters to non-core-collapsed Milky Way
clusters with the same range of central surface brightness,
we find the median $r_0$ to be 0.77~pc for the M31 clusters and
1.14~pc for the Milky Way clusters.
The difference is almost certainly due to selection effects:
most of the targeted HST observations
were of bright M31 clusters with high central surface brightness,
and larger low-surface-brightness clusters 
(the `Palomar'-type Milky Way clusters are the extreme examples)
would have been very difficult to detect in M31.
The lowest surface brightness objects we expect
to detect in our median exposure have ${\mu}_V(0)\approx 20$.

Except for the lack of confirmed  core-collapsed clusters in M31, discussed 
in the previous subsection, the range of concentration parameters
is similar for the M31 and Milky Way clusters. The median
values of $c$ are 1.40 for the non-core-collapsed
Milky Way clusters and 1.43 
for the M31 clusters. Six of our M31 clusters fall into 
a region of parameter space where no Milky Way clusters are found:
$c\lesssim 1.1$, $r_0<1$~pc (these objects are also the outliers in the
$r_0, r_h$ plot). The identities of all are questionable: 132--000 had 
been previously been classified as a star from its spectrum \citep{b00}, 
268--000 is an unconfirmed C-class cluster from \citet{bat87} with few known properties, 
000--M91 is an unconfirmed
candidate first discovered by \citet{m98}, and the other three objects 
are new cluster candidates. All are faint and small; several have rather 
poor fits to the King models. Three are projected very close to the center
of M31. Their estimated relaxation times
at the half-mass radius range from $2\times 10^7$ to $2.5\times 10^8$~yr.
The relaxation times are quite short compared to the Hubble time, and if the objects projected
near the center of M31 are truly near the nucleus, they should have been destroyed long ago.
We are uncertain about the nature of these objects, and suggest that higher
resolution images and/or spectra may be needed to fully understand them.
To confirm that the M31 clusters are well-characterized by King models,
we performed a principal component analysis with the five
structural parameters $M_V, {\mu}_V(0), r_0, r_h,$ and $c$. We find, as did 
\citet{dm94} for the Milky Way clusters,
that the dimensionality of this dataset is $D=3$.
The three-parameter King model is adequate to describe the 
surface brightness profiles of M31 globular clusters.

\subsection{Galactocentric distance}

\begin{figure}
\includegraphics*[scale=0.7]{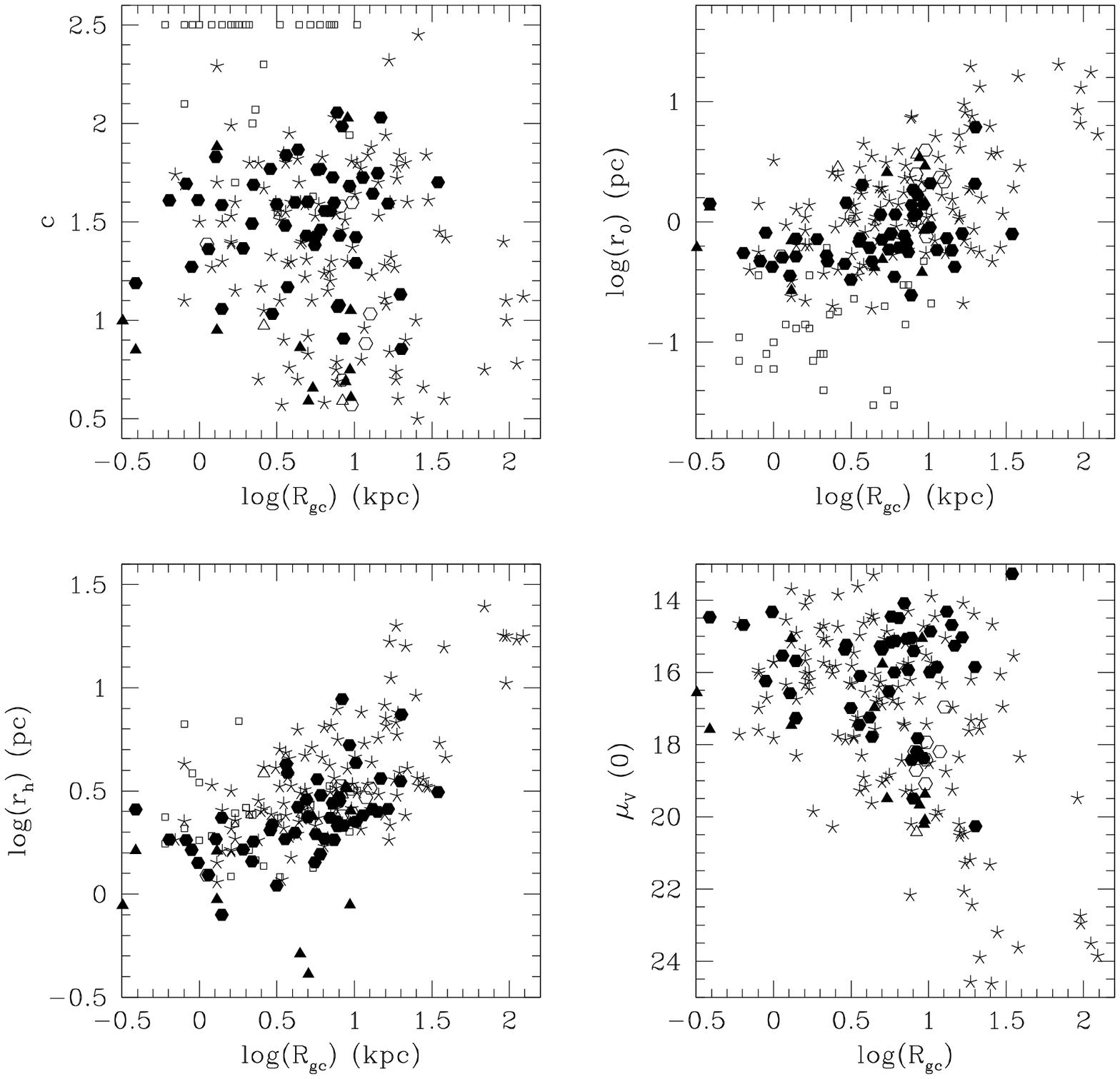}
\caption
{Structural parameters of M31 and Milky Way globular 
clusters as a function of galactocentric distance
(true three-dimensional distance for Milky Way clusters
and projected distance on the sky for M31 clusters).
Symbols as in Figure~\ref{ellip_plot}.
\label{rgc_plot}}
\end{figure}

Figure~\ref{rgc_plot} shows the Milky Way and M31 structural parameters
as a function of galactocentric distance $R_{gc}$. There is no
clear correlation of ${\mu}_V(0)$ with $R_{gc}$, except for
the tendency, mentioned above, for the very low surface brightness
Milky Way GCs to be located far from the galaxy center. This is
unsurprising, since such clusters would be easily destroyed
by dynamical forces nearer to the galaxy center.
Also expected from dynamical considerations is that core-collapsed clusters 
should have smaller average $R_{gc}$ (since the stronger tidal field accelerates
the clusters' dynamical evolution); Figure~\ref{rgc_plot} shows that
this is true for the Milky Way.
Most of the M31 core-collapse candidates are within 2~kpc of the center of M31,
and a KS test shows the $R_{gc}$ distributions of the
core-collapse candidates and the rest of the sample
to be different at the 95\% confidence level.
There is no significant trend of $c$ with $R_{gc}$ for the {\em non}-core
collapsed clusters in either M31 or the Milky Way.
Both $r_h$ and $r_0$ are correlated with galactocentric distance, which
has been noticed before for Milky Way clusters \citep*{vmp91}. These
authors suggest that, while large clusters with small $R_{gc}$ could
have been destroyed, there is no equivalent reason for the lack of small
clusters at large $R_{gc}$, so the correlation is due to physical
conditions at the time of cluster formation. The data in Figure~\ref{rgc_plot}
suggest that similar conditions affected the formation of the M31 globular clusters.

\subsection{Ellipticities}

\begin{figure}
\includegraphics*[scale=0.7]{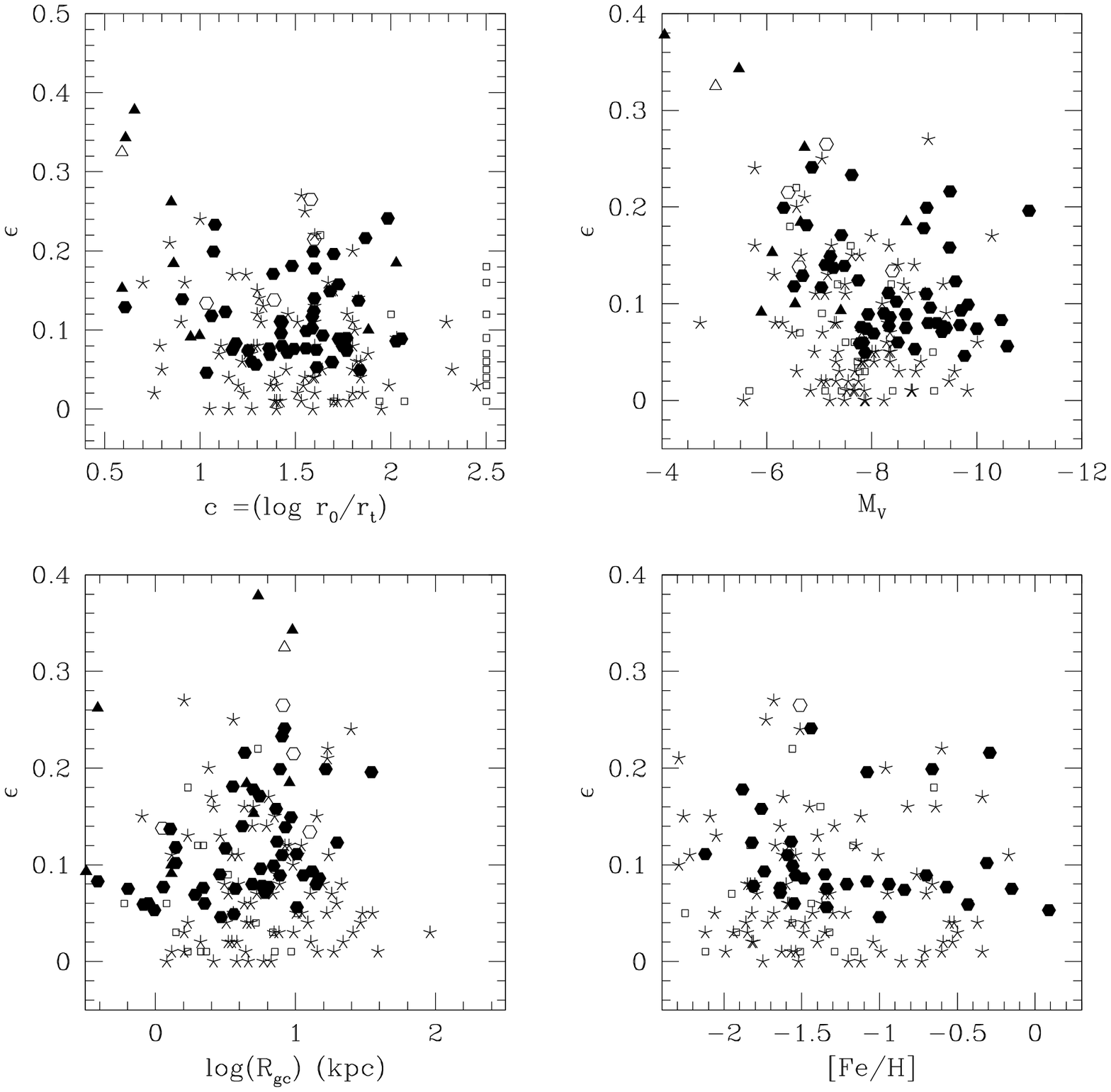}
\caption
{Ellipticity versus other properties of M31 and Milky Way
globular clusters.
Hexagons are previously-known M31 globulars and triangles are new objects.
Filled symbols are likely globular clusters;
outlined symbols are blue clusters which may not be old GCs.
Stars are non-core-collapsed Milky Way GCs; small squares are core-collapsed
Milky Way GCs (for which $c$ is set arbitrarily to 2.5).
\label{ellip_plot}}
\end{figure}

Cluster rotation, rather than tidal forces, is the generally accepted 
explanation for cluster flattening. A general picture, summarized by \citet{dp90},
is that GCs form with some angular momentum and are initially flattened by rotation.
As escaping stars carry away angular momentum and mass, clusters rotate more
slowly and become rounder. The rotation model makes several predictions.
One is that more compact clusters, which evolve more quickly, should
be rounder. \citet{ws87} found this to be the case for Milky Way clusters.
We plot ellipticity against several other parameters for both sets of clusters 
in Figure~\ref{ellip_plot}. We find no clear relation between $c$ and $\epsilon$ for 
Milky Way clusters, but we do see that low-concentration clusters in M31 are generally 
more elliptical, as predicted. 
A second prediction is that clusters with larger velocity dispersions should 
be rounder because they rotate more slowly,
due to conservation of angular momentum in the sum
of internal velocity dispersion and rotation \citep{ssg96}. These authors find
a relation between $\epsilon$ and ${\sigma}_v$ in M31 clusters in their data,
but when we add later velocity dispersion measurements by \citet{djo97} to
\citeauthor{ssg96}'s ellipticity data, we see no obvious correlation. 

Conflicting claims have been made about correlations of globular cluster
ellipticities with other properties. \citet{lup89} claimed that ellipticity was 
anti-correlated with metallicity for both Milky Way clusters and 
his sample of 18 M31 clusters; \citet{ssg96} and \citet{ws87} found no such 
correlation. Our data in Figure~\ref{ellip_plot} show little correlation
(only about half of our M31 clusters have measured metallicities), but
if we bin the data in [Fe/H] we find that the most metal-poor objects are slightly 
more elliptical. \citet{dp90} claimed a relation between luminosity and ellipticity
for both M31 and Milky Way clusters, with the brightest clusters 
being the roundest; \citet{ssg96} found the same for M31.%
\footnote{The brightest 
clusters in each galaxy, $\omega$~Cen and G1, are both quite flattened, 
with $\epsilon\sim0.2$. There have been suggestions that neither object is
a true globular cluster \citep{hr00,mey01}, so their failure to follow this trend may not
be meaningful.\label{foot-g1}} 
\citet{lup89} did not find a luminosity-ellipticity
relationship in his sample. Our data again show no clear correlation, but with binned data 
we do find that the least-luminous clusters tend to be more elliptical. 
\citet{ws87} found that the most elliptical
Milky Way clusters were found near the galactic plane but did not claim
a correlation of $\epsilon$ with $R_{gc}$; \citet{ssg96} and \citet{lup89}
also found no such correlation in their M31 cluster samples.
We find a slight trend in the opposite direction: the innermost clusters are 
slightly less elliptical.
\citet{dm94} found no correlations of Milky Way GC ellipticity with any other properties,
and suggest that this may be because of the difficulties in measuring
ellipticity: the effect and its measurement errors are of comparable size.
Our results suggest that there may be more subtle effects in the M31 globular
cluster system which \citeauthor{dm94} did not find in the MW system.

\subsection{Metallicities}

\begin{figure}
\includegraphics*[scale=0.7]{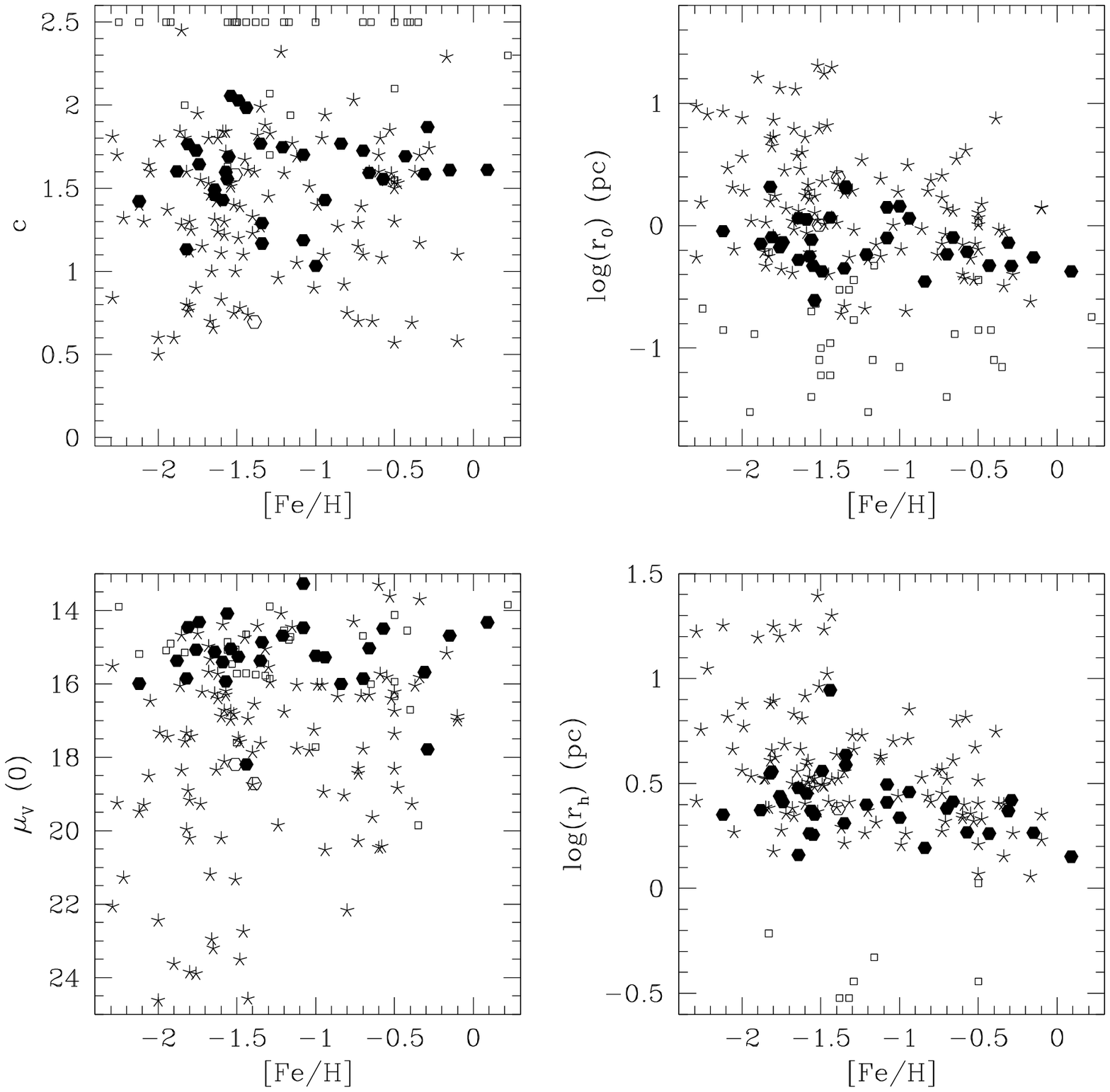}
\caption
{Metallicity versus structural parameters of M31 and Milky Way 
globular clusters. Symbols as in Figure~\ref{ellip_plot}.
\label{feh_plot}}
\end{figure}

\begin{figure}
\includegraphics*[scale=0.7]{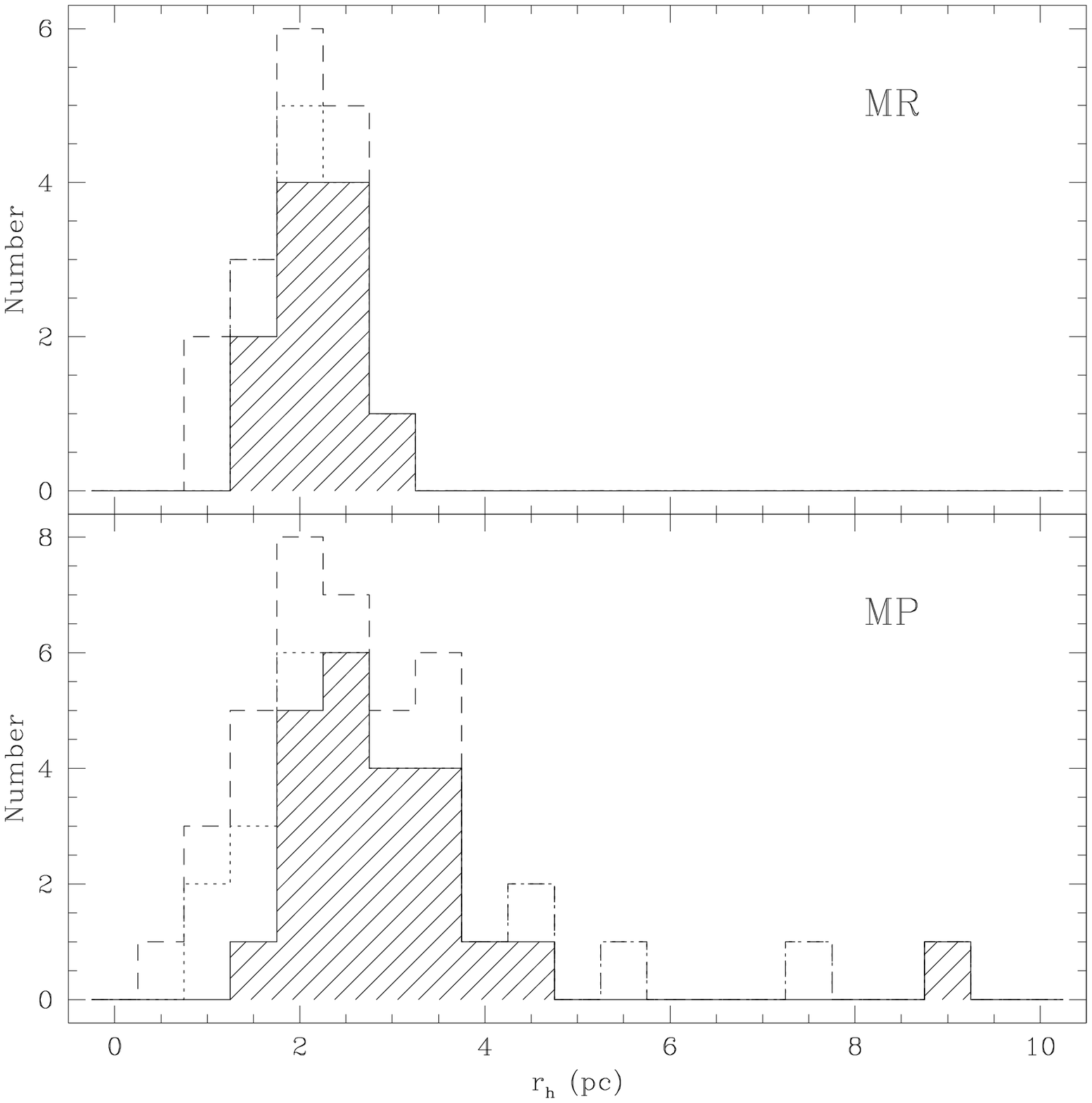}
\caption
{Size distribution of M31 globular clusters in two metallicity groups.
Solid line/shaded histogram includes only clusters with spectroscopic metallicities.
Dotted line histogram also includes clusters with color-derived metallicities from
\citet{b00}. Dashed line histogram also includes clusters with rough
metallicity indicators from single colors.
\label{feh_size}}
\end{figure}

As noted above, there is some weak evidence for lower-metallicity
M31 clusters to be more elliptical. Figure~\ref{feh_plot} shows that 
there is essentially no correlation of metallicity with concentration
or central surface brightness, but there does appear to be a
correlation with size, as measured by $r_0$ or $r_h$.
This has been noticed before: examining globular cluster systems
in many galaxies, both \citet{kws99} and \citet{lar01} found
that the metal-rich clusters had slightly smaller average values of $r_h$.
We assigned M31 clusters to metallicity groups based on either spectroscopic metallicities,
color-derived metallicities from \citet{b00}, or the HST $V-I$ color 
\citep[the criterion used by][]{lar01}.
Figure~\ref{feh_size} shows size distribution for the two groups. The median sizes for the
two groups are similar to values found for other galaxies:
2.17~pc for the metal-rich clusters and 2.76~pc for the metal-poor ones.
Some of the size difference could be due to $R_{gc}$: metal-rich clusters are 
more likely to be near the center of M31, and we have already shown that there 
is a gradient of  $r_h$ with $R_{gc}$. 
There is still a size difference between the two metallicity groups
for clusters with $R_{gc}>2$~kpc;
however, a KS test does not show the $r_h$ differences to be statistically 
significant, and the small number of metal-rich clusters (17 in total and only 11 with
$R_{gc}>2$~kpc) makes our conclusions uncertain. 
The correspondence with the results for the Milky Way and other
galaxies is certainly suggestive. The size difference could indicate the different
pericenter distances of the clusters' orbits, as might be expected from their
different kinematics.

\subsection{Parameter Correlations and the Fundamental Plane}

\begin{figure}
\includegraphics*[scale=0.7]{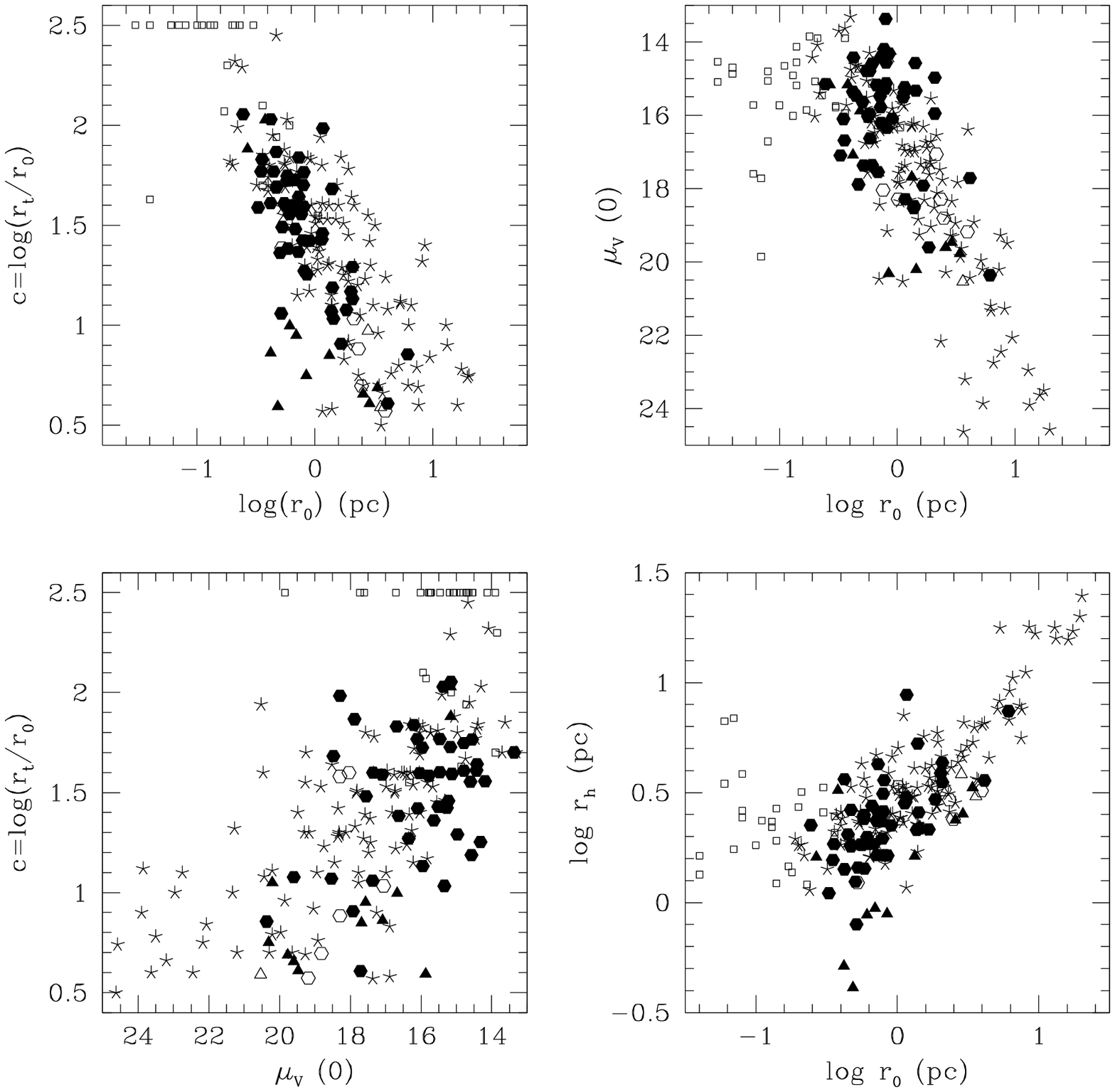}
\caption
{King model structural parameters for M31 and MW GCs.
Symbols as in Figure~\ref{ellip_plot}.
\label{coreplots}}
\end{figure}

Globular cluster structure is described by four parameters:
concentration $c$, scale radius $r_0$, central surface brightness ${\mu}_V(0)$,
and central mass-to-light ratio ${\Upsilon}_0$ or velocity dispersion
${\sigma}_0$. Figure~\ref{coreplots} shows that there are strong correlations between 
the King model parameters of both Milky Way and M31 globular clusters,
meaning that clusters do not inhabit the full four-dimensional parameter space.
\citet{dj95} found a pair of bivariate correlations in Milky Way GC
parameters which imply the existence of a `globular cluster fundamental plane'.
Only about 20 M31 GCs have measurements of mass-to-light ratios \citep{dg97,djo97}, 
but they appear to fall on the same plane as the Milky Way clusters. 
\citet{bel98} posited the existence of a `fundamental straight line' as expected
if GCs represent a family of objects with constant core mass evolving toward 
core collapse. \citet{mcl00} examined the Milky Way GC fundamental plane
in detail and disputed \citeauthor{bel98}'s interpretation, pointing out that
cluster cores cannot be viewed as dynamically distinct entities since 
they do not obey the virial theorem.

Any linearly independent combination of the parameters described above
is a complete basis for describing GC structure, and \citet{mcl00}
chose the set $c$, ${\Upsilon}_0$, luminosity $L$, and binding energy $E_b$
to describe the Milky Way clusters. He showed that the Milky Way clusters'
fundamental plane was described by the equations 
\begin{equation}\label{mlconst}
{\Upsilon}_{V,0} =1.45\pm0.1
\end{equation}
and
\begin{equation}\label{fp}
E_b=A L^{\gamma} \Rightarrow \log E_b  = (39.89\pm0.38) + (2.05\pm0.08)\log L
\end{equation}
Although equation~\ref{mlconst} can be used in computing
$E_b$ for the Milky Way clusters, this does {\em not} imply
an automatic correlation between $E_b$ and $L$: the two
parameters are linearly independent.

Since our HST observations provide no new information about M31 clusters' 
mass-to-light ratios, we cannot directly test \citeauthor{mcl00}'s result
of a constant ${\Upsilon}_0$ for M31 clusters. However, \citeauthor{mcl00}
used the fundamental plane to predict the existence of several
`monovariate' correlations which involve only the King model parameters
(his equations A14 and A17). These are:
\begin{equation}
\log r_0 = 41.2208 - \log A + 2\log {\Upsilon}_0 + (2-\gamma) \log L - f(c)
\end{equation}
and
\begin{equation}
{\mu}_V(0) = 232.466 - 5\log A + 10\log {\Upsilon}_0 + (7.5-5\gamma)\log L - g(c)
\end{equation}
where $A$ and $\gamma$ are given above, and $f$ and $g$ are `nonhomology terms'
\citep[for details, see][]{mcl00}. We can use these predictions
to see whether the M31 
clusters' properties are compatible with the Milky Way fundamental plane.

\begin{figure}
\includegraphics*[scale=0.7]{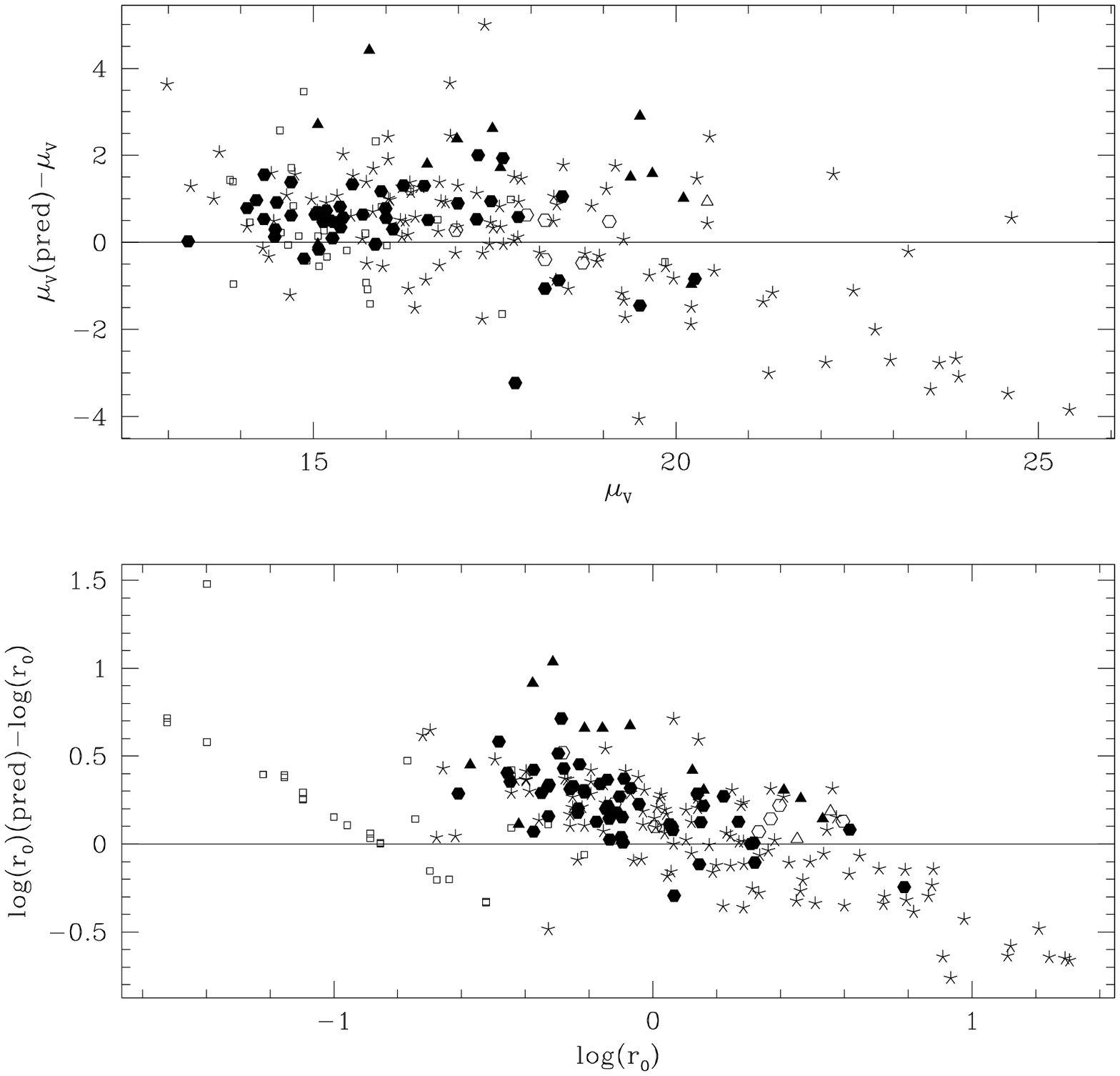}
\caption
{Difference between predicted \citep[from the fundamental plane relations in][]{mcl00}
and measured $r_0$ and ${\mu}_V(0)$ for M31 and MW GCs. The Milky Way constant 
mass-to-light ratio ${\Upsilon}_0=1.45$ was used to compute the predicted values.
The apparent trend for the Milky Way core-collapsed clusters (small squares) 
is not meaningful since these objects do not have a core radius.
Symbols as in Figure~\ref{ellip_plot}. 
\label{monovar}}
\end{figure}

Figure~\ref{monovar} plots the difference between the 
fundamental plane predictions and the measured values of
$r_0$ and ${\mu}_V(0)$ for M31 and Milky Way GCs. Table~\ref{tbl-fp}
gives the statistics of the differences between predicted 
and measured values (core-collapsed Milky Way clusters are not
included). The standard errors of the means are similar 
for the two sets of clusters, which is
is somewhat surprising, given that we expect the observational
errors to be larger for the M31 data. It implies that the two
sets of clusters have about the same amount of scatter 
about the fundamental plane. The mean values
are consistent with zero for the Milky Way clusters, but offset
from zero for the M31 clusters. This could imply 
that the M31 clusters have a different mass-to-light ratio
${\Upsilon}_0$ and/or $E_b-L$ intercept $A$ from the Milky Way clusters, 
or that there are systematic errors in our M31
cluster measurements.

\begin{figure}
\includegraphics*[scale=0.7]{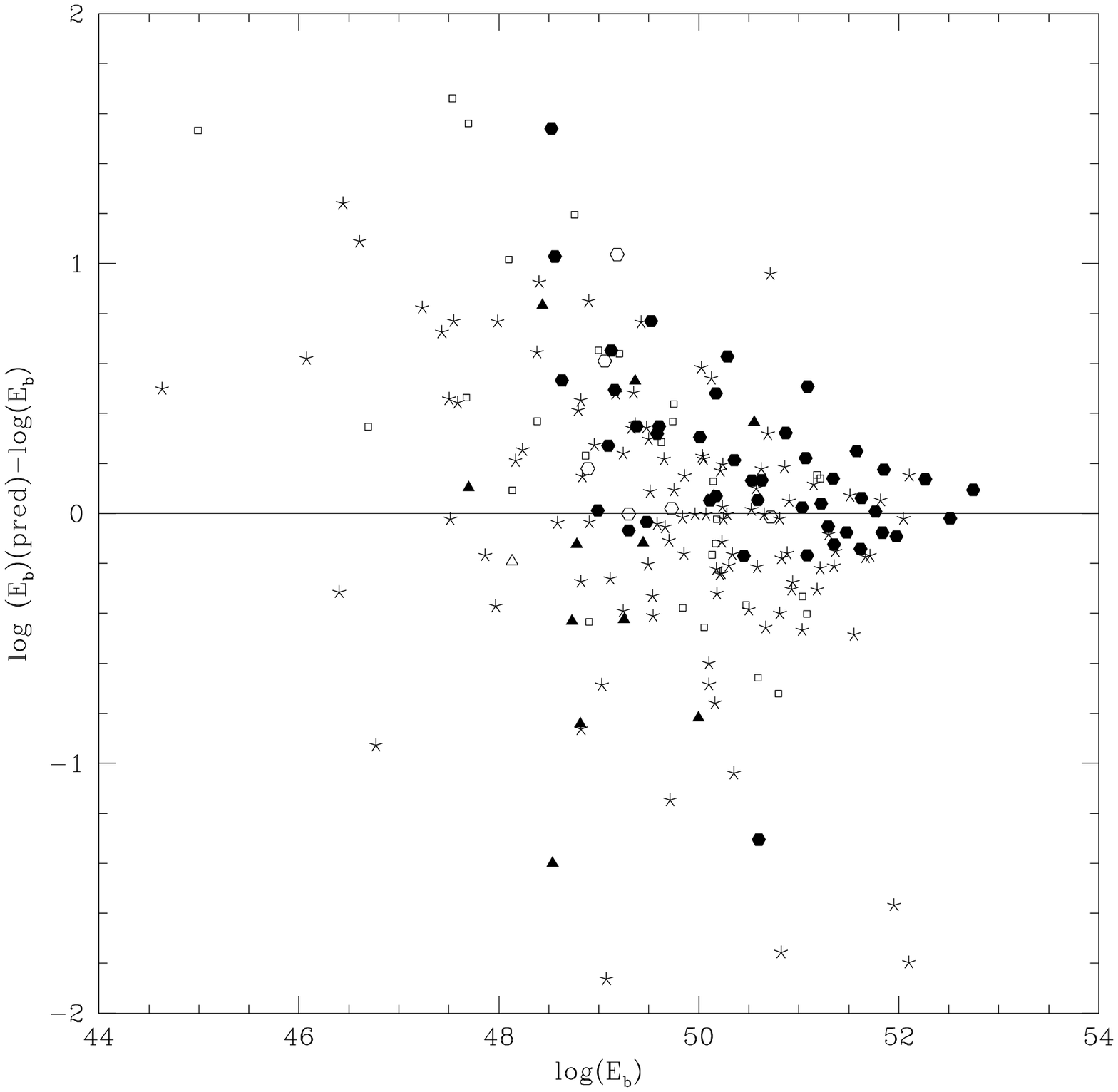}
\caption
{Difference between predicted \citep[from the fundamental plane relations in][]{mcl00}
and measured $E_b$ for M31 and MW GCs.
Symbols as in Figure~\ref{ellip_plot}. 
\label{funplane}}
\end{figure}

We can test this idea by examining the fundamental plane 
predictions of $E_b$ as a function of $L$ for the M31 clusters. 
The difference between $E_b$ predicted from $L$ (using equation~\ref{fp}) 
and the measured value 
(computed from \citeauthor{mcl00}'s equation~5c)
also has a term of the form $a = 2\log{\Upsilon}_0-\log A$, but with the
opposite sign to the $r_0$ and ${\mu}_V(0)$ differences.
Figure~\ref{funplane} shows the difference between the 
fundamental plane predictions of $E_b(L)$ and the measured values 
for M31 and Milky Way GCs. Again the Milky Way clusters' mean
difference is consistent with zero. In this case the M31 clusters have a
slightly larger scatter than the Milky Way clusters, as well
as an offset. $\Delta \log E_b$ has the same sign as 
$\Delta r_0$ and $\Delta {\mu}_V(0)$, which means that 
a different value of $a$ for the M31 clusters cannot simultaneously 
account for all three offsets. 

Can the differences between fundamental plane predictions and
M31 GC observations be explained by a combination of true
differences between the M31 and Milky Way GC fundamental planes
and systematic errors in the observations? Yes.
Increasing our measured M31 $\log (r_0)$ values by 0.12
and decreasing $a$ by 0.09, while leaving ${\mu}_V(0)$ unchanged,
resulted in all $\Delta$ values consistent with zero. 
Subtracting 0.57~mag from our measured M31 ${\mu}_V(0)$ values
and decreasing $a$ by 0.20, while leaving $r_0$ unchanged,
gave the same result. The dependence of the $\Delta$ values on $c$ 
(through the nonhomology terms $f(c)$ and 
$g(c)$ above and ${\cal E}(c)$ in $E_b$)
is not as straightforward as that on $\log(r_0)$ and ${\mu}_V(0)$, 
but experiment showed that
increasing the measured values of $c$ for M31 clusters by 0.25 
(with no changes in $r_0$ or $a$)
resulted in $\Delta$ values consistent with zero. 
Other combinations of parameter changes
might also result in a better fit to the fundamental
plane, but we concentrate here on the simplest ones.

The systematic changes which improve the M31 clusters'
fit to the Milky Way GC fundamental plane also
result in a slightly better match with the Milky Way clusters
in Figure~\ref{coreplots}. This is not surprising, since
the correlations shown there are due to the existence
of the fundamental plane. But are the changes reasonable?
A change in $a$ of 0.09, if interpreted as due to 
the M31 clusters' mass-to-light ratio, implies 
${\Upsilon}_{0,{\rm M31}}=1.3$, well within the measured values 
for M31 clusters. 
Increasing the measured $r_0$ by 30\% ($\Delta \log (r_0)=+0.12$)
would worsen the 
agreement of our measurements with those by other groups,
although it would make the median $r_0$ for M31 clusters
closer to that for Milky Way clusters.
A systematic error in $r_0$ would seem more likely
to result in our measured values being too large
(because of the limited spatial resolution) rather than too small.
Subtracting 0.56 from our measured ${\mu}_V(0)$ marginally improves
the agreement with other groups, and it is plausible that 
the limited spatial resolution could have resulted in our measuring
central intensities which were 60\% of the true value.
Increasing $c$ by 0.25 (and thereby increasing $r_t$ by $10^{0.25}$) 
marginally worsens agreement of $r_t$ with other values, but
also seems a plausible effect of limited spatial resolution.

More information is needed to distinguish between the 
various possible reasons for the offset between 
fundamental plane predictions and M31 cluster observations.
The small scatter of the mean M31 offsets implies that the
M31 clusters may well have a constant mass-to-light ratio, but additional
measurements would be very useful to confirm this and
to determine the value of ${\Upsilon}_0$. Additional measurements
of King model parameters, both from the existing HST data and
from future data, will help to clarify whether there are
systematic errors in our method. Even with the offsets, it is clear 
that M31 and Milky Way  clusters have a limited and very similar 
range of properties, controlled by strong correlations. 
If additional galaxies' GCs have similar fundamental planes,
this will strengthen the case for a `universal' GC formation mechanism, in which
GC properties are controlled by very few parameters 
\citep[possibly the initial protocluster gas mass:][]{mcl00,bel98}.

\section{Summary}

We use HST images of M31 globular clusters to measure the clusters' sizes, 
shapes, and best-fit King model
parameters. Cluster departures from sphericity are consistent with being caused by 
rotation, although there are also indications of relations between ellipticity 
and luminosity and metallicity. We find a slight difference between the
half-light radii of metal-rich and metal-poor clusters, consistent with 
previous results on clusters in other galaxies. The M31 clusters are
well-described by the three-parameter family of King models. They have
approximately the same range of parameter values as the Milky Way clusters, 
except that there are few faint, low-concentration clusters in our M31 sample
due to observational selection effects.
The scatter about the fundamental plane relations is
very similar for Milky Way and M31 clusters, although the M31 clusters
are offset from the Milky Way cluster relations.
This effect may be due to an intrinsic difference in the
two galaxies' clusters' fundamental planes or to systematic errors
in our measurements: further information is needed.
The overall similarity of the two fundamental planes 
implies that the formation and evolution of GCs must have been very similar 
in the two galaxies. 

\acknowledgments
We thank J. Grindlay, G. Harris, W. Harris and S. Zepf for helpful discussions, 
and K. Stanek and R. Di Stefano for critically reading the manuscript.
We also thank the referee, whose report was helpful in clarifying several issues.

\begin{figure}
\includegraphics*[scale=0.7]{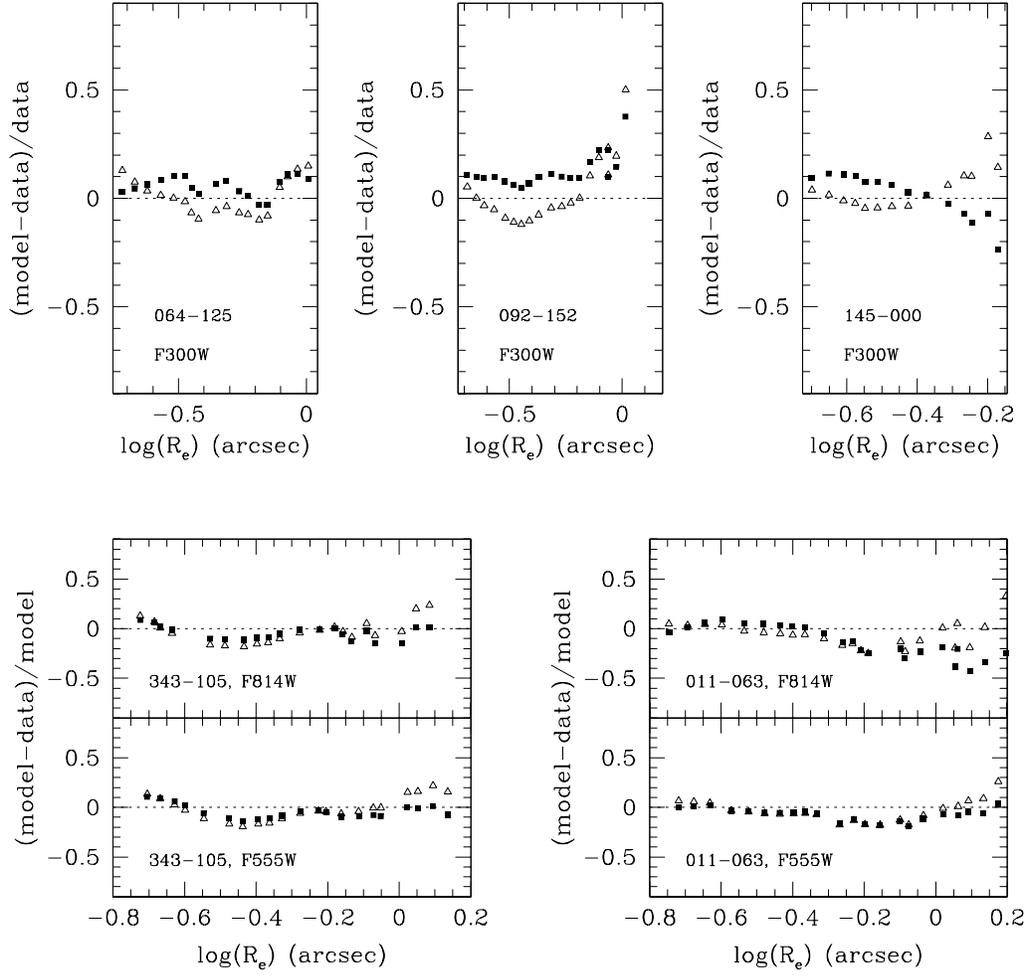}
\caption
{Difference between modeled and measured
surface brightness profiles for possible core-collapsed M31 GCs.
Solid squares are King models $-$ data; open triangles 
are power-law surface brightness profiles $-$ data.
\label{fig-corecoll}}
\end{figure}

\begin{figure}
\includegraphics*[scale=0.7]{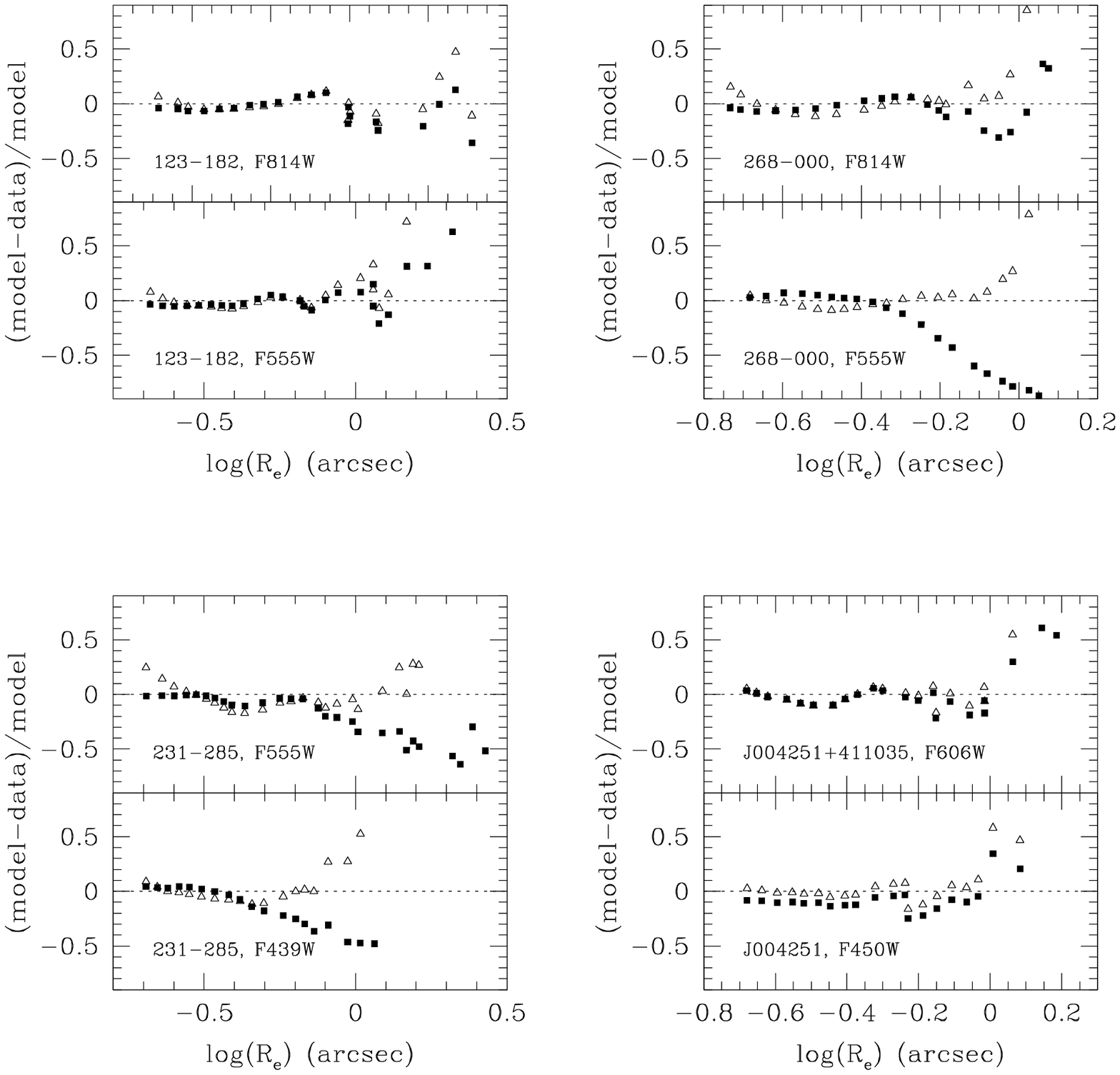}
\figurenum{5}
\caption
{Continued.}
\end{figure}

\begin{deluxetable}{lllllll}
\tablewidth{0pt}\
\tablecaption{Shape parameters for GCs in M31 HST fields\label{tbl-param}}
\tablehead{\colhead{name}&\colhead{ellipticity\tablenotemark{a}}&\colhead{PA\tablenotemark{b}}&\colhead{$r_0$(\arcsec)\tablenotemark{c}}&\colhead{$r_t$(\arcsec)}&\colhead{$c$}
&\colhead{${\mu}_V(0)$\tablenotemark{d}}}
\startdata
000--001 &$0.20\pm0.01$&$121\pm 1$  &  0.21& 10.50& 1.70&13.65\\         
000--D38 &$0.34\pm0.04$&$108\pm 1$  &  0.77&  3.11& 0.61&19.72\\         
000--M091&0&	0		    &  1.09&  4.42& 0.61&20.56\\         
006--058 &$0.08\pm0.02$&$ 74\pm 2$  &  0.16&  5.76& 1.56&15.00\\         
009--061 &$0.12$&$ 31$		    &  0.15&  5.86& 1.60&15.23 ($I$)\\   
011--063 &$0.09\pm0.02$&$ 76\pm 6$  &  c0.06&  7.28& 2.06&15.40\\        
012--064 &$0.08\pm0.01$&$ 46\pm 9$  &  0.21& 12.33& 1.77&14.91\\         
018--071 &$0.15\pm0.04$&$  4\pm 2$  &  0.37& 17.42& 1.68&18.73\\         
020D--089&$0.14$&$ 21$		    &  0.16&  6.36& 1.60&16.18 ($I$)\\   
027--087 &$0.07\pm0.02$&$ 97\pm29$  &  0.30&  8.72& 1.46&16.63\\         
030--091 &$0.10\pm0.01$&$118\pm17$  &  0.21&  5.50& 1.42&17.32\\         
045--108 &$0.08\pm0.01$&$ 40\pm 1$  &  0.30&  8.14& 1.43&15.94\\         
058--119 &$0.10\pm0.00$&$138\pm 1$  &  0.20&  7.29& 1.56&14.56\\         
064--125 &$0.06$&$ 53$              &  c0.12&  6.03& 1.69&17.36 (F300W)\\ 
068--130 &$0.22\pm0.02$&$ 42\pm 1$  &  0.12&  9.04& 1.87&19.28\\          
070--133 &0&	0                   &  (edge) &\nodata &\nodata &\nodata\\
076--138 &$0.09\pm0.01$&$ 69\pm 1$  &  0.12&  6.92& 1.77&16.12\\          
077--139 &$0.14\pm0.02$&$132\pm10$  &  (edge) &\nodata &\nodata &\nodata\\
092--152 &$0.08$&$109$              &  c0.14&  4.28& 1.49&18.07 (F300W)\\
097D--000&0&	0                   &  0.75&  6.83& 0.97&18.91 ($I$)\\
101--164 &$0.07$&$ 80$              &  0.19&  4.39& 1.37&17.97 (F300W)\\  
109--170 &$0.10\pm0.03$&$ 72\pm 9$  &  0.19&  7.44& 1.59&16.03\\          
110--172 &$0.05$&$ 49$		    &  0.38&  4.09& 1.03&15.83\\          
114--175 &$0.06\pm0.01$&$132\pm4$   &  0.21&  3.99& 1.27&16.84\\          
115--177 &$0.08\pm0.02$&$ 63\pm 7$  &  0.14&  5.92& 1.61&15.04\\          
123--182 &$0.14\pm0.04$&$ 62\pm 6$  &  c0.09&  6.34& 1.83&16.93\\         
124--NB10&$0.07\pm0.01$&$164\pm 3$  &  0.22&  3.89& 1.25&14.56\\          
127--185 &$0.08\pm0.01$&$ 64\pm 4$  &  0.37&  5.74& 1.19&15.04\\          
128--187 &$0.08\pm0.02$&$176\pm 1$  &  0.13&  3.49& 1.36&15.25 ($I$)\\    
132--000 &$0.09\pm0.02$&$ 40\pm59$  &  0.16&  1.59& 1.00&15.63 ($I$)\\    
134--190 &$0.16\pm0.00$&$113\pm 1$  &  (edge) &\nodata &\nodata &\nodata\\ 
143--198 &$0.05\pm0.01$&$158\pm 7$  &  0.11&  4.53& 1.61&14.83\\          
145--000 &$0.14$&$ 92$	  	    &  c0.14&  3.36& 1.39&19.43 (F300W)\\ 
146--000 &$0.06$&$151$	  	    &  0.12&  6.12& 1.69&18.19 (F300W)\\  
148--200 &$0.07$&$ 27$	            &  (edge) &\nodata &\nodata &\nodata\\
153--000 &$0.05$&$ 70$	            &  (edge) &\nodata &\nodata &\nodata\\
155--210 &$0.12\pm0.01$&$ 80\pm15$  &  0.09&  3.49& 1.59&17.49\\
156--211 &$0.05\pm0.02$&$ 67\pm 4$  &  0.19& 14.37& 1.84&16.45\\
160--214 &$0.18\pm0.00$&$  2\pm 1$  &  0.18&  5.44& 1.48&17.80\\
167--000 &$0.04$&$ 97$              &  (edge) &\nodata &\nodata &\nodata\\
205--256 &$0.08\pm0.04$&$152\pm22$  &  0.53&  9.12& 1.17&17.14 (F300W)\\
231--285 &$0.17\pm0.02$&$136\pm33$  &  c0.15&  3.96& 1.38&16.88\\
232--286 &$0.18\pm0.01$&$ 42\pm 1$  &  0.19&  7.56& 1.60&15.69\\ 
233--287 &$0.11\pm0.02$&$ 74\pm 8$  &  0.30&  7.99& 1.43&15.88\\ 
234--290 &$0.07\pm0.01$&$ 71\pm16$  &  0.09&  5.33& 1.77&16.35\\ 
240--302 &$0.16\pm0.01$&$ 98\pm 1$  &  0.18&  9.35& 1.73&15.42\\ 
264--NB10&$0.26\pm0.04$&$142\pm17$  &  0.35&  2.54& 0.85&17.93\\ 
268--000 &$0.12\pm0.04$&$103\pm50$  &  c0.14&  1.56& 1.06&17.62\\
279--D68 &$0.20\pm0.09$&$ 79\pm77$  &  0.36&  4.39& 1.07&18.78\\ 
311--033 &$0.09\pm0.01$&$ 54\pm 7$  &  0.19&  8.46& 1.64&15.13\\ 
315--038 &$0.13\pm0.02$&$159\pm11$  &  0.57&  6.56& 1.03&17.31\\ 
317--041 &$0.11\pm0.02$&$ 66\pm23$  &  0.24&  6.31& 1.42&16.50\\ 
318--042 &$0.19\pm0.03$&$ 70\pm 5$  &  (edge) &\nodata &\nodata &\nodata\\
319--044 &0&	0                   &  0.62&  4.73& 0.88&18.54\\ 
324--051 &0&	0 		    &  0.66&  3.19& 0.69&19.06\\ 
328--054 &$0.27\pm0.05$&$159\pm 5$  &  0.27& 10.70& 1.58&18.54\\ 
330--056 &$0.14\pm0.01$&$102\pm 8$  &  0.44&  3.62& 0.91&18.17\\ 
331--057 &$0.24\pm0.06$&$ 70\pm 3$  &  0.31& 29.53& 1.98&18.54\\ 
333--000 &$0.23\pm0.02$&$ 26\pm17$  &  0.49&  5.83& 1.08&19.85\\ 
338--076 &$0.06\pm0.01$&$102\pm34$  &  0.55& 10.74& 1.29&15.34\\ 
343--105 &$0.09\pm0.01$&$ 70\pm22$  &  c0.11& 11.91& 2.03&15.58\\
358--219 &$0.12\pm0.02$&$ 63\pm 5$  &  0.55&  7.40& 1.13&16.18\\ 
368--293 &0&	0 		    &  1.04&  3.86& 0.57&19.43\\ 
374--306 &$0.21\pm0.02$&$106\pm 1$  &  0.20&  7.90& 1.60&18.29\\ 
379--312 &$0.09\pm0.02$&$ 55\pm 3$  &  0.15&  8.16& 1.73&16.27\\ 
384--319 &$0.20\pm0.01$&$121\pm 1$  &  0.21&  8.27& 1.59&15.50\\ 
386--322 &$0.08\pm0.01$&$140\pm 3$  &  0.15&  8.54& 1.75&15.05\\ 
468--000 &0&	0 		    &  1.61& 11.55& 0.86&20.61\\ 
NB39     &$0.13\pm0.04$&$ 28\pm17$               &  0.22&  1.26& 0.75&17.96\\               
M31GC~J004304+412028  &$0.09\pm0.04$& $75\pm3$	 &  0.18&  1.65&0.95 &17.82\\   
M31GC~J004251+411035  &$0.10\pm0.01$& $116\pm20$ &  c0.07&  5.45&1.88 &16.61\\  
M31GC~J004258+405645  &$0.18$&$175$		 &  0.11&  0.80&0.86 &17.33\\   
M31GC~J004301+405418  &$0.15\pm0.04$&$ 30\pm14$	 &  0.13&  0.50&0.59 &18.04\\   
M31GC~J004312+405303  &$0.38\pm0.01$&$ 44\pm 4$	 &  0.68&  3.06&0.66 &21.46\\   
M31GC~J004103+403458  &$0.38\pm0.14$&$ 87\pm 8$	 &  (edge) &\nodata &\nodata &\nodata\\
M31GC~J004537+413644  &0&	0		 &  0.90&  4.39&0.69 &21.04\\   
M31GC~J004030+404530  &0.19&123		 &  0.10& 10.66&2.03 &15.31\\   
M31GC~J004027+414225  &$0.32\pm0.06$&$128\pm 4$	 &  0.95&  3.70&0.59 &20.78\\   
M31GC~J004051+404039  &0&	0		 &  0.38&  4.26&1.05 &20.45\\   
\enddata
\tablenotetext{a}{Ellipticity is defined as $\epsilon=1-(b/a)$, where $a$ and $b$ are the lengths
of the semi-major and semi-minor axes, respectively}
\tablenotetext{b}{Position angle is measured in degrees east from north}
\tablenotetext{c}{`c' indicates core-collapse candidates.}
\tablenotetext{d}{Bandpass names indicate central surface brightness measured
in other than $V$.}
\end{deluxetable}

\begin{deluxetable}{lrrrrc}
\tablewidth{0pt}
\tablecaption{Structural parameters for cluster 000--001 (G1)\label{tbl-g1}}
\tablehead{\colhead{Source}&\colhead{$r_0$}&\colhead{$r_h$}&\colhead{$r_t$}
&\colhead{$c$}&\colhead{${\mu}_V(0)$}\\
\colhead{}&\colhead{(\arcsec)}&\colhead{(\arcsec)}&\colhead{(\arcsec)}&\colhead{}&\colhead{(mag arcsec$^{-2}$)}}
\startdata
this work & 0.21 & 0.82 & 10.5 & 1.70 & 13.65 \\
Rich et~al. (1996) & 0.17 & 0.70 & 28.2 & 2.22 & 13.5\\
Meylan et~al. (2001) & 0.14 & 3.7 & 54 & 2.59 & 13.47\\
\enddata
\end{deluxetable}

\begin{deluxetable}{lrrrr}
\tablewidth{0pt}
\tablecaption{Fundamental plane predictions $-$ measurements\label{tbl-fp}}
\tablehead{\colhead{}&\colhead{$\Delta \log(r_0)$}&\colhead{$\Delta {\mu}_V(0)$}&\colhead{$\Delta E_b$}&\colhead{$N$}}
\startdata
Milky Way & $0.03\pm 0.03$\tablenotemark{a} & $0.15\pm0.15$ & $-0.03\pm0.05$ & 110\\
M31       & $0.20\pm 0.03$ & $0.45\pm0.14$ & $0.25\pm0.09$ & 45 \\
\enddata
\tablenotetext{a}{Values are mean $\pm \sigma / \sqrt{n}$.} 
\end{deluxetable}

\end{document}